\documentclass[11pt]{article}

\usepackage[utf8]{inputenc}
\usepackage[T1]{fontenc}
\usepackage{amsmath, amssymb, amsthm}
\usepackage{graphicx}
\usepackage{hyperref}
\usepackage{cleveref}

\usepackage[margin=1in]{geometry}

\title{Standards for trustworthy AI in the European Union: technical rationale, structural
challenges, and an implementation path}

\author{
  Piercosma Bisconti\\
  Department of Computer, Control and Management Engineering\\
  Sapienza University of Rome\\
  \texttt{piercosma.bisconti@uniroma1.it}
    \and 
  Marcello Galisai\\
  Department of Computer, Control and Management Engineering\\
  Sapienza University of Rome\\
  \texttt{galisai.2084612@studenti.uniroma1.it}
}

\date{\today}

\begin{document}

\maketitle

\begin{abstract}
This white paper examines the technical foundations of European AI standardization under the AI Act. It explains how harmonized standards enable the presumption of conformity mechanism, describes the CEN/CENELEC standardization process, and analyzes why AI poses unique standardization challenges including stochastic behavior, data dependencies, immature evaluation practices, and lifecycle dynamics. The paper argues that AI systems are typically components within larger sociotechnical systems, requiring a layered approach where horizontal standards define process obligations and evidence structures while sectoral profiles specify domain-specific thresholds and acceptance criteria. It proposes a workable scheme based on risk management, reproducible technical checks redefined as stability of measured properties, structured documentation, comprehensive logging, and assurance cases that evolve over the system lifecycle. The paper demonstrates that despite methodological difficulties, technical standards remain essential for translating legal obligations into auditable engineering practice and enabling scalable conformity assessment across providers, assessors, and enforcement authorities.
\end{abstract}

\section{Introduction}

Trustworthy AI is not a slogan. In the European Union it is an operational objective translated into binding obligations through the AI Act and implemented by a dense infrastructure of standardization, conformity assessment, and market surveillance.\cite{EUAIAct2024, EURegulation1025} To move from abstract values to engineering practice, Europe relies on technical standards that describe requirements, test methods, and documentation artefacts in a manner that can be audited and reproduced. This white paper presents a technical view that is deliberately concrete. It explains what standards are in the European context and why they matter due to their presumption of conformity effect, describes how CEN and CENELEC convert policy intent into normative text, addresses the methodological difficulty of standardizing AI where mature and shared methods are still emerging, and examines the additional complexity that arises because AI is frequently a component inside a larger sociotechnical system. It then argues that progress is both possible and necessary, outlining how risk management and reproducible technical checks can be framed so that providers and assessors can obtain stable evidence of conformity, even under uncertainty about models and data.

\section{Standards in the EU and the presumption of conformity mechanism}

European standards are consensus based documents that specify technical requirements and supporting methods to meet the essential requirements of Union law. The legal effect does not come from the standard itself but from its citation in the Official Journal as a harmonized standard. When a provider follows a cited standard, it can claim a presumption of conformity with the corresponding legal requirements.\cite{EURegulation1025} This presumption does not eliminate the obligation to comply, but it shifts the evidentiary burden. Instead of defending a proprietary approach line by line, the provider references a shared method whose adequacy has been evaluated during the harmonization process. In the AI domain this is decisive because the legal obligations are formulated as outcomes and processes rather than design recipes. Without standards, every provider would need to invent its own proof of adequacy, and every assessor would need to evaluate idiosyncratic evidence with limited comparability across cases.

CEN and CENELEC are the European organizations that develop these standards.\cite{CENCENELECJTC21, JRCAIStandards2024} They operate a structured workflow that begins with a mandate or standardization request from the European Commission. The request identifies which legal provisions require support from standards, sets target dates, and indicates the expected type of deliverable.\cite{EuropeanCommission2023Mandate} After acceptance, the technical work proceeds in Joint Technical Committee 21 and its working groups. Drafts are prepared under national delegation. Normative text is constructed using a controlled vocabulary of requirements and recommendations. Verbs are chosen carefully because they encode obligations. Shall indicates a requirement, should indicates a recommendation, may indicates permission, and can indicates a possibility. Each clause is classified as normative or informative to keep the presumption clear. Annexes serve specific functions. The Z annex maps requirements in the standard to specific legal provisions and is scrutinized by independent consultants appointed by the Commission. If the mapping is sound, the final text advances through enquiry and formal vote in the national standards bodies. Once adopted and cited, it becomes part of the legal infrastructure that supports conformity assessment, market surveillance, and judicial review.

Two further elements of the European approach are worth noting. First, alignment with global standards is encouraged through the Vienna Agreement which enables parallel work with ISO and IEC.\cite{ViennaAgreement, FrankfurtAgreement} This reduces divergence between European and global practice and helps European industry operate internationally. Second, the New Legislative Framework provides a modular set of conformity routes that range from internal control to full quality assurance with third party involvement.\cite{EURegulation765,EUDecision768} Standards are written in a way that allows their requirements and test methods to be used across these routes. The outcome is a coherent system where law sets the essential results to be achieved, standards specify reproducible ways to demonstrate that those results have been achieved, and conformity assessment verifies and documents the demonstration.

\section{Why standards for AI are technically difficult today
}

Standardization assumes the existence of stable methods that can be specified, applied, and checked by independent parties who obtain consistent results. AI unsettles this assumption. Many AI techniques are stochastic. The same model with the same inputs and nominally identical configuration can produce different outputs across runs.\cite{Henderson2018DeepRLReproducibility, Starace2025PaperBench} Deterministic inference is possible by fixing seeds and controlling hardware and library versions, yet this control often fails to replicate production conditions that involve distributed systems, dynamic loads, and model updates. Reproducibility therefore requires more than a recipe. It requires a definition of acceptable variation and a way to measure whether variation is within limits that preserve safety and performance.

Data add another layer of difficulty. The behaviour of a model is inextricable from the data used for training, fine tuning, and evaluation. Data provenance, representativeness, and label quality all affect downstream behaviour. Yet in many cases providers do not own the entire data pipeline. They rely on data collected by third parties, or on open data whose documentation is incomplete. Standards that require absolute control by the provider over all data would block legitimate development. Standards that ignore data governance would fail to ensure trustworthy behaviour. The only workable path is a risk based approach where the level of control and evidence is calibrated to the impact of the system and to the risk of data induced failures.

Evaluation practices are also immature.\cite{Semmelrock2025ReproducibilityML} Benchmarks often reward surface pattern matching rather than robust generalization. Small changes in dataset distribution can cause large drops in performance.\cite{Hardt2024BetterBench} Socially salient properties such as bias and potential for manipulation are context dependent and require scenario based testing rather than a single score. For large language models and other generative systems, harmful behaviour can emerge from complex interactions with prompts and tools. This makes point estimates fragile and calls for families of tests, stressors, and red team procedures that explore the boundaries of behaviour. Turning such families into standards that are precise enough for conformity assessment is inherently difficult. The test must be defined in a way that independent labs can execute with similar outcomes, while leaving space for rapidly evolving techniques and threat models.

Lifecycle dynamics complicate the picture. Models are updated frequently. Data drift changes the operating environment. Feedback driven fine tuning alters behaviour over time. A one off evaluation at product launch is not sufficient. Standards need to articulate post market monitoring duties, logging and traceability requirements, and triggers for re evaluation. At the same time, the cost of repeated full evaluation can be prohibitive. The standard must therefore define what can be monitored continuously through telemetry, what requires periodic structured testing, and what changes require a formal reassessment because they affect intended purpose, risk profile, or performance envelopes.

Finally, there is the question of interpretability. For many AI systems we cannot provide mechanistic explanations that are accepted as engineering proof of safety. The evidence must be built from empirical tests, statistical assurance, and structured arguments that connect claims to data and to tests. Assurance cases are suitable for this purpose, yet they need a template and a vocabulary that allow assessors to understand and challenge the logic. Building such templates into standards is possible, but it takes time and iterative validation across many products and sectors.

\section{AI as a component and the limits of horizontal standardization
}

AI is rarely a standalone product. In most deployments it is a component embedded in a larger system that includes hardware, software, human operators, and organizational processes. The behaviour of that larger system depends on the interactions among components. An AI based predictor that estimates the probability of equipment failure has a limited direct effect. The actual risk arises from how operators interpret the estimate, how maintenance schedules adapt to warnings, how false alarms affect trust, and how workflows change over time. In a clinical context an algorithm that supports diagnosis interacts with patient pathways, consent processes, clinical governance, and liability frameworks. In a financial context a fraud detection model interacts with customer service and legal redress. The technical and social environment define what counts as safe and acceptable behaviour.

Horizontal standards that address AI in general therefore face a structural constraint. They can require risk management, documentation, data governance, robustness evaluation, transparency measures, and human oversight capabilities. They can define classes of functions such as classification, prediction, or generation, and define associated properties to be measured. They cannot define performance thresholds that are meaningful across all sectors. The tolerance for error, delay, or variance is determined by domain specific risk, regulatory context, and societal expectations. A miss rate that is acceptable in content moderation would be unacceptable in a medical diagnosis support tool. A logging policy that is adequate for a retail chatbot would not satisfy the needs of a forensic audit in a law enforcement system.

The solution is architectural rather than rhetorical. Horizontal standards provide the scaffold. They specify method families, process obligations, and evidence structures that are applicable across AI systems. Sectoral or application profile standards then refine the general methods for particular classes of use. A profile for clinical decision support would reference the horizontal risk management and testing methods but would add domain specific metrics, operating conditions, and thresholds aligned with existing health technology assessment practices. A profile for automated driving support would connect horizontal AI robustness methods with standards for functional safety and roadworthiness already in force. The same applies to banking, critical infrastructure, education, and public administration. In this layered model, general AI standards avoid over specification while still creating a common language and a shared toolkit for development and assessment.

This layered structure has two implications for conformity assessment. First, the evidence portfolio for an AI system must be assembled from both horizontal and sectoral sources. Providers need to show that they followed the general methods and that they met domain specific targets. Second, notified bodies and market surveillance authorities must be able to interpret both layers, which argues for cross training and for cooperation between authorities responsible for the sector and authorities responsible for digital technologies. Without this cooperation there is a risk of gaps where each authority assumes that the other is responsible for specific controls.

\section{Why technical standards remain necessary and how to make them work
}

Despite the methodological and structural challenges, technical standards are necessary for trustworthy AI. A risk management requirement that lives only in law remains a general principle. To be workable, it must be decomposed into tasks, artefacts, and checks that can be executed and audited. A robustness requirement must be associated with test suites, fault models, and acceptance criteria that are documented and reproducible. A transparency requirement must result in user facing notices, system cards, and logs that contain specific fields and follow controlled vocabularies so that they can be parsed, aggregated, and compared. The function of a standard is to translate each of these requirements into concrete obligations that a provider can implement inside a quality management system.

A practical approach begins with the lifecycle. The standard asks the provider to declare intended purpose, operating conditions, and unacceptable misuse that is reasonably foreseeable. It requires a hazard analysis and risk estimation that covers failures of function, data induced failures, interaction induced failures, and adversarial misuse. It requires traceability from hazards to controls and from controls to verification and validation activities. It requires documentation of the data pipeline, including provenance, preprocessing, and bias analysis. It requires model documentation that records architectures, training regimes, hyperparameters, and tuning strategies. It requires evaluation plans that describe test datasets, metrics, and statistical analysis. It requires performance statements that include uncertainty and conditions of validity. It requires deployment controls such as access management, monitoring, incident response, and fallback procedures. It requires post market monitoring plans that define telemetry, thresholds for investigation, and triggers for corrective action.

These instructions are not novel in engineering. What is novel is the way they must be adapted to the peculiarities of modern AI. Reproducibility cannot be equated with byte identical outputs.\cite{Desai2025Reproducibility} It must be defined as stability of measured properties under controlled variation. Standards can define reproducibility protocols that fix seeds and versions for the baseline measurement, then measure sensitivity under controlled perturbations that reflect production conditions. They can require that safety relevant metrics be accompanied by confidence intervals and that labs report the variance they observe across runs and across hardware configurations. They can require that providers supply evaluation containers or environments that allow third parties to rerun tests with minimal friction while documenting any deviations from the original environment.

Risk management must account for the fact that hazard likelihoods are difficult to estimate for novel models. Standards can require scenario based analysis that builds from known classes of failure such as distribution shift, prompt induced undesired actions, tool misuse, and reward hacking. They can require the use of stressor libraries that simulate realistic threats, and they can require periodic refresh of these libraries to account for new attack surfaces. They can also require adversarial testing by teams that are organizationally independent from development, with reporting that feeds into corrective action processes. Where thresholds are not known, standards can require staged release with progressive exposure and guardrails, coupled with monitoring that detects early signs of harmful drift.

Transparency obligations are meaningful if they yield artifacts that can be read and checked. Standards can specify the content and format of technical documentation that accompanies an AI system, including sections for data governance, model description, training and evaluation procedures, performance claims, risk controls, and change history. They can define system cards that present a concise public summary of intended use, limitations, and known risks. They can specify interaction notices that inform users about AI mediated functionality, including residual risks and escalation paths to human support. For oversight to be real, the human must have actionable information and tools. Standards can require interfaces that allow operators to inspect key decision variables, override system outputs, and record decisions for audit.

Logging and traceability deserve special attention. In AI, incidents and near misses often become visible only after forensic analysis of logs. Standards can require immutable logs for safety relevant functions, with time synchronization, identity management, and tamper detection. They can define log schemas that include inputs, outputs, confidence or calibration information, model identity and version, data sources used at inference time, and any tool calls or external actions. For remote biometric identification, for example, logs must allow reconstruction of the operational cycle, the datasets used for enrolment, the thresholds applied, and the identity of the human verifiers who validated a match. For generative systems that can trigger external actions through tools, logs must record the tool chain, parameters, and outputs so that a misfire can be diagnosed and contained.

Conformity assessment benefits from evidence that is standardized in structure and content. A provider should be able to assemble a portfolio where each claim is backed by a test report or a process record that follows agreed templates. The assessor should be able to navigate this portfolio and verify cross references without relying on the provider to interpret idiosyncratic documentation. Standards can therefore define the skeleton of the portfolio, including indexes, cross reference matrices mapping evidence to legal provisions, and change logs that track updates after deployment. When third party assessment is required, this structure reduces assessment time and increases consistency across assessors.\cite{EuropeanCommissionNLFEvaluation}

The relationship between horizontal standards and sectoral profiles must be managed deliberately. Horizontal standards should avoid embedding sector specific thresholds. They should instead define hooks that profiles can use to introduce domain targets and acceptance criteria. For example, a horizontal robustness standard can define families of tests for out of distribution resilience, calibration, and failure under stress. A clinical profile can reference those tests and specify that given the use in triage the model must maintain minimum sensitivity and specificity at defined prevalence ranges, and that confidence estimates must remain calibrated across specified demographic groups. The same base tests can support a rail transport profile that defines entirely different thresholds and operating conditions. This approach preserves a single method family and enables reuse of tooling and lab expertise while allowing sectors to enforce their specific safety envelopes.

Parallel alignment with global standards avoids duplication and fragmentation. Many of the process obligations described above are already present in management system standards for AI, risk management frameworks, and quality management systems. Where possible, European standards should import or reference global clauses rather than rewrite them.\cite{ISO42001} This not only reduces conflict for companies operating globally but also accelerates the time to market for supporting tools and services. At the same time, where European law introduces specific obligations that have no direct global counterpart, European standards must be precise. The mapping from standard requirements to legal provisions in the Z annex should be traceable at clause level. This mapping is a contract with the presumption of conformity.\cite{Pouget2025AIActStandards} If it is vague the presumption becomes weak and litigation risk increases. If it is precise providers and assessors have a stable target and market surveillance authorities can enforce consistently.

\section{Toward a workable scheme of risk management and reproducible checks
}

A realistic scheme for trustworthy AI in Europe can be articulated as follows. The provider operates an AI quality system that integrates the standard requirements into development and operations.\cite{CENCENELEC2025prEN18286} The system is anchored by a risk management process that tracks hazards from conception to retirement. For each hazard class the provider defines controls, identifies verification and validation activities, and assigns responsibility. Verification focuses on the product. Validation focuses on the fit between product and intended purpose. Both are supported by test methods defined in standards and by test data whose provenance and quality are documented. The provider builds an assurance case that connects claims to evidence and states residual risk. The case is updated when models are retrained or retuned, when data pipelines change, or when operating conditions shift. The provider operates monitoring that detects departures from expected behaviour, records incidents, and triggers corrective action. Significant changes trigger re evaluation according to a decision rule that is defined in advance and documented. Throughout, the provider keeps technical documentation that follows the structure defined in the standard and maintains logs that allow auditors and investigators to reconstruct relevant events.

To make this scheme reproducible, standards must require sufficient detail in the definition of tests and artefacts. For example, a standard might require calibration curves for probabilistic outputs along with the method used to compute calibration and the bins and temperature scaling parameters used. It might require error decomposition that separates data noise from model bias and from evaluation variance. It might require stress testing under controlled perturbations such as input noise, prompt variations, and composition with simple tools. It might require seed sets for inference and a declaration of non deterministic components so that labs understand what variance to expect. It might require that providers publish a statement of evaluation scope that lists which functions are covered and which interactions are not yet tested, with a plan and schedule for coverage expansion. None of these requirements solve AI safety in the abstract. They make it possible to evaluate and compare, to detect regressions, and to hold providers accountable for the claims they make.

\section{Conclusion}
European standards for trustworthy AI are not a luxury. They are the only credible way to transform legal obligations into engineering practice that can be executed by providers, audited by assessors, and enforced by authorities. The presumption of conformity mechanism gives standards a special role. It allows providers to rely on shared methods instead of bespoke arguments and it allows the legal system to scale assessment and enforcement in a complex technological landscape. The difficulty is real. AI lacks universally accepted methods for many aspects of evaluation and assurance. Models change quickly and behave differently across contexts. AI is almost always embedded in systems where human behaviour, organizational processes, and sector specific rules shape outcomes. Horizontal standards cannot fix domain thresholds, and sectoral standards cannot redefine the fundamentals of AI risk management and testing.
The answer is a layered and disciplined approach. Horizontal standards define lifecycle obligations, documentation structures, risk management processes, families of test methods, logging and oversight capabilities, and portfolio structures for conformity. Sectoral profiles specify operating conditions and acceptance criteria that reflect domain risk and regulation. Both layers remain aligned with global work to the extent possible. Reproducibility is reconceived as stability of measured properties within declared variance bounds, supported by controlled environments and sensitivity analysis. Evaluation is diversified beyond single benchmarks to include adversarial testing, scenario based analysis, and monitoring in deployment. Evidence is organized in assurance cases that accumulate over the lifecycle rather than static documents produced at launch.
This is demanding, but it is feasible. Europe has the institutional machinery to make it work. CEN and CENELEC can convert legal intent into precise clauses, test methods, and mappings that support presumption of conformity. Providers can build quality systems that implement the obligations as part of their development and operations. Notified bodies can evaluate evidence against stable criteria. Market surveillance can enforce based on logs and documentation that follow defined structures. The result is not a guarantee of perfect behaviour. It is a framework that makes behaviour governable and evidence accountable, which is the essence of trustworthiness in engineered systems.

\bibliographystyle{plain}
\bibliography{references}


\end{document}